# *Pre-exposure prophylaxis* (PrEP) versus *treatment-as-prevention* (TasP) for the control of HIV: Where does the balance lie?

Brian G. Williams,* Eleanor Gouws,† John Hargrove,* Cari van Schalkwyk,* and Hilmarie Brand*

\* South African Centre for Epidemiological Modelling and Analysis, Stellenbosch, South Africa
† Joint United Nations Programme on HIV/AIDS (UNAIDS), Geneva, Switzerland

Correspondence to BrianGerardWilliams@gmail.com

**Abstract**

Anti-retroviral drugs can reduce the infectiousness of people living with HIV by about 96%—'treatment as prevention' or TasP—and can reduce the risk of being infected by an HIV positive person by about 70%—'pre-exposure prophylaxis' or PrEP—raising the prospect of using anti-retroviral drugs to stop the epidemic of HIV. The question as to which is more effective, more affordable and more cost effective, and under what conditions, continues to be debated in the scientific literature. Here we compare TasP and PreP in order to determine the conditions under which each strategy is favourable. This analysis suggests that where the incidence of HIV is less than 5% or the risk-reduction under PrEP is less than 50%, TasP is favoured over PrEP; otherwise PrEP is favoured over TasP. The potential for using PreP should therefore be restricted to those among whom the annual incidence of HIV is greater than 5% and TasP reduces transmission by more than 50%. PrEP should be considered for commercial sex workers, young women aged about 20 to 25 years, men-who-have-sex with men, or intravenous drug users, but only where the incidence of HIV is high.

## Introduction

The XIX International AIDS Conference held in Washington DC in 2012, was dominated by the prospect of ending the epidemic of HIV,[1] and bringing about what Secretary of State Hilary Clinton called an 'AIDS Free Generation'.[2] The discussions were motivated by the realization that anti-retroviral therapy (ART) keeps people alive, renders them much less infectious, makes HIV-negative people much less likely to be infected, and stops vertical transmission. A still unresolved question concerns the relative impact and benefits of using ART to render HIV-positive people uninfectious—Treatment as Prevention or TasP—as compared to using ART to protect HIV-negative people from becoming infected—pre-exposure prophylaxis or PrEP.[3-20]

In 2012 the International Association of Physicians for AIDS Care published a consensus statement on the use of TasP and PrEP[21] but did not consider in detail the conditions under which TasP should be favoured over PrEP and *vice versa*. For the purposes of this discussion we assume that:

- Under TasP, people would be tested at regular intervals, start ART as soon as they are found to be HIV positive, and then maintained on ART for life.
- Under PrEP, people would start ART as soon as they are thought to be at risk of HIV. If they remain HIV-negative they would stay on ART until they are no longer at risk; if they become infected with HIV they would be switched to a different regimen, if necessary, and kept on ART for life.

In both cases people would be visited at regular intervals: infected people would need to receive drugs and be monitored for compliance, viral rebound and drug resistance; uninfected people would need to receive drugs and be monitored for compliance under PrEP and tested for HIV under TasP. We will assume that requirements for support and testing under the two approaches are equivalent. The important comparison is therefore between the number of person years on ART after they become infected under TasP and the number of person years on ART both before they become infected but remain at risk and after they become infected under PrEP.

Here we only consider the individual benefit and do not include the long term impact on transmission. This analysis is therefore relevant to a situation in which we have a small group of people, or even an individual person who can be managed either under TasP or PrEP.

## Methods

Let the annual incidence of HIV in the population under consideration be $\alpha$. We assume 1) that PrEP reduces the risk that the person becomes infected by a factor of $\rho$ so that the incidence among those on PrEP is $\beta = \rho\alpha$; 2) that they remain at risk for a time $d$ after which they are no longer at risk where $d$ could refer to the time for which women remain in sex-work or to the time for which people remain sexually active and at risk; 3) that the life expectancy of uninfected people and those on ART is $l$.

The two strategies are illustrated schematically in Figure 1. The blue line shows people who are receiving drugs under TasP. Those in areas B, C, D and E are on ART because they are HIV-positive; those in areas D and E are no longer at risk but are HIV-positive, having been infected earlier. The red line shows proportion of people who are receiving drugs under PrEP. Those in area A and B are receiving drugs while they are HIV-negative; those in area C have become HIV-positive, in spite of being on PrEP; those in area E have acquired HIV but are no longer at risk and must be maintained on ART until they die.

From Figure 1 it is clear that the difference in the number of person years on ART under a policy of TasP as opposed to PrEP is determined by the relative sizes of areas A and D since the other areas are common to both strategies.

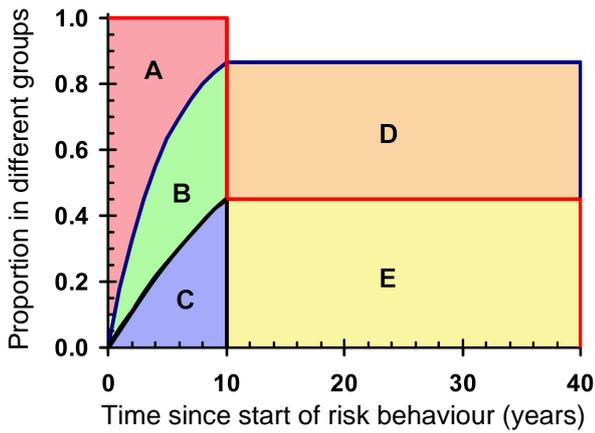

Figure 1. Proportion of people on ART under TasP (blue line) and PrEP (red line). A: people on PrEP only; B and C: people on TasP and PrEP; D: people on TasP only; E people on PrEP and TasP. Here the duration of risk behaviour is $d = 10$ years and life-expectancy is $l = 40$ years after the start of risk behaviour.

The incidence of HIV without PrEP is $\alpha$ and with PrEP be $\beta$ so that the relative risk of infection with PrEP is $\rho = \beta/\alpha$. Then the areas in Figure 1 are:

$$A = \frac{1}{\alpha}\left(1 - e^{-\alpha d}\right) \quad\quad 1$$

$$A + B = \frac{1}{\beta}\left(1 - e^{-\beta d}\right) \quad\quad 2$$

$$C = d - \frac{1}{\beta}\left(1 - e^{-\beta d}\right) \quad\quad 3$$

$$D = (l - d)\left(e^{-\beta d} - e^{-\alpha d}\right) \quad\quad 4$$

$$E = (l - d)\left(1 - e^{-\beta d}\right) \quad\quad 5$$

The total number of person years on ART under PrEP is A+B+C+E and under TasP is B+C+D+E.

## Results

We assume here that people are at risk of HIV for 10 years and will live for 40 years after they are exposed to the risk of HIV infection. This would correspond, for example, to a situation in which young people are at risk from the age of 20 years to the age of 30 years and live to the age of 60 years. Alternatively, one might consider a women engaging in sex-work from the age of 20 to 30 years and then stopping but continuing to live to the age of 60 years.

In Figure 2 the number of person years on ART under PrEP and under TasP is plotted as a function of the annual incidence of HIV without PrEP (horizontal axis) and the relative risk of infection if a person is on PrEP as compared to those that are not (vertical axis). The heavy black line shows the combinations of values where PrEP and TasP are equivalent. Above the black line TasP would lead to having fewer person years on ART; below the black line PrEP would lead to having fewer years on ART. If the annual incidence is about 0.3%, say, indicated by the junction between the two green areas, then PrEP would lead to having 8 times as many person years on ART. If PrEP reduces incidence by less than 40% (relative risk > 0.6; peak of heavy black line in Figure 1) PrEP would always lead to an increase in the number of person years on ART. If the incidence of infection is less than 0.3% per year (intersection of heavy black line and horizontal axis) TasP would always be favoured over PreP. The best prospects for using PrEP would be when the annual incidence is about 20% (peak of the heavy black line in Figure 2). Even then PrEP would have to reduce the risk of HIV infection by 90% in order to halve the number of person years on ART (line separating pink and red areas in Figure 2.)

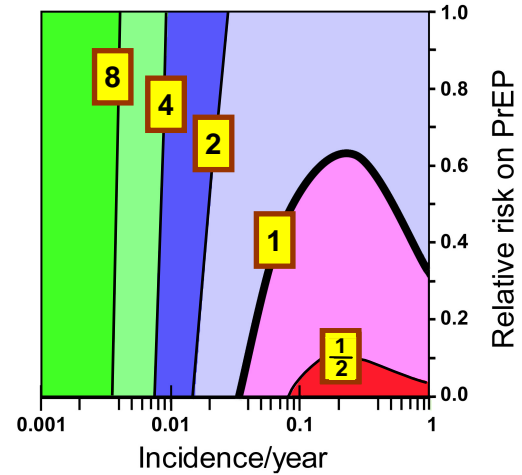

Figure 2. Area plot of the ratio, $R$, of the number of person years on ART under PrEP compared to TasP. The horizontal axis gives the annual incidence of HIV without PrEP on a logarithmic scale; the vertical axis gives the relative risk of HIV infection for people on PrEP as compared to those that are not. Numbers in boxes give the value of $R$ along each contour. Below the heavy black line PrEP is favoured; above the black line TasP is favoured.

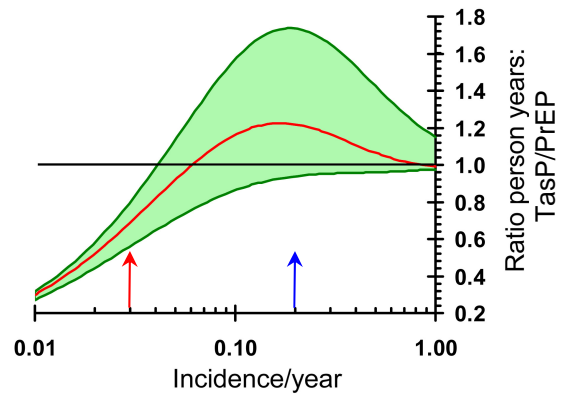

Figure 3. The ratio of the number of person years on ART under PrEP compared to TasP (red line) with 95% confidence limits (green lines). Here PrEP reduces the risk of infection by 62% (0.16%–78%).[4] Red arrow: the incidence of infection in the control arm of PrEP trial in Botswana[4] of 3.1%/year; blue arrow: nominal incidence of 20%/year.

We use this model to compare PrEP against TasP based on the results of a study carried out in Botswana[4] in which PrEP reduced the risk of infection by 62% (0.16%–78%). As shown in Figure 3, the incidence rate reported in the control arm of the trial was 3.1% per annum (red arrow) so that TasP rather than PrEP would result in about 30% fewer person-years on ART. If the annual incidence were much higher at 20% per annum, say, (blue arrow) the data from the trial shows that TasP without PrEP would lead to 24% more person years on ART which would favour PrEP but it should be noted that the 95% confidence interval for this estimate is −4% to 75%) so that the estimate is not significantly different from zero ($p > 0.05$).

## Discussion

This analysis suggests that PrEP is unlikely to lead to a reduction in the number of person years requiring ART unless the incidence of ART without PrEP is greater than about 5% per year and PrEP reduces the risk of infection by at least 50%.

Here we represent the risk as a step-function of time; since we have good data on the age-specific incidence of infection from many epidemics we could generalize the argument to apply in such cases. Note that the balance between the two approaches will depend on the duration of risk, here taken to be 10 years, and it will be important to investigate further the way in which this affects the results.

Although the adult incidence of infection in nearly all countries of the world is less than 5% per year, and often very much less, incidence rates of more than 10% have been reported among young women in southern Africa[22,23] and among intravenous-drug user and commercial sex workers. In such situations PrEP could have an important role to play, provided the reduction in the risk of infection is greater than about 50%. The PrEP study in Botswana reported low rates of retention partly because of nausea, vomiting and dizziness brought on by the use of ART and if these problems can be overcome it may be possible to get a higher retention rate under PrEP.

We have excluded the impact of treatment on transmission. Under either approach widespread treatment would reduce incidence and this would swing the balance further in favour of TasP and against PrEP.